\documentclass[a4paper,11pt]{article}
\usepackage{pos}
\usepackage{amsmath,graphicx,multirow,tabularx,physics}
\usepackage{slashed,listings}
\usepackage{float}
\usepackage{wrapfig}

\definecolor{airforceblue}{rgb}{0.36,0.54,0.66}
\definecolor{burgundy}{rgb}{0.5,0.0,0.13}
\definecolor{blue-violet}{rgb}{0.54,0.17,0.89}
\title{A machine learning approach to the classification of phase transitions in many flavor QCD}
\ShortTitle{A ML approach to many flavor QCD}

\author[a]{F. Karsch}
\author[a]{A. Lahiri}
\author*[a]{M. Neumann}
\author[a]{C. Schmidt}

\affiliation[a]{Fakultät für Physik, Universität Bielefeld,\\
Universitätsstraße 25, Bielefeld, Germany}

\emailAdd{karsch@physik.uni-bielefeld.de}
\emailAdd{alahiri@physik.uni-bielefeld.de}
\emailAdd{neumann@physik.uni-bielefeld.de}
\emailAdd{schmidt@physik.uni-bielefeld.de}

\abstract{Normalizing flows are generative machine learning models which can efficiently approximate probability distributions, using only given samples of a distribution. This architecture is used to interpolate the chiral condensate obtained from QCD simulations with five degenerate quark flavors in the HISQ action. From this a model for the probability distribution of the chiral condensate as function of lattice volume, quark mass and gauge coupling is obtained. Using the model, first order and crossover regions can be classified and the boundary between these regions can be marked by a critical mass. An extension of this model to studies of phase transitions in QCD with variable number of flavors is expected to be possible.}

\FullConference{%
The 39th International Symposium on Lattice Field Theory,\\
8th-13th August, 2022,\\
Rheinische Friedrich-Wilhelms-Universität Bonn, Bonn, Germany
}

\newcommand{\pbp}{\langle \bar{\psi}\psi \rangle}

\begin{document}
\maketitle

\section{Introduction}

Almost 40 years ago Pisarski and Wilczek argued, that the chiral phase transition for vanishing masses and three or more flavors ($N_f$) should be of first order~\cite{Pisarski1984}. 
To this day, the search for the orderness of the transition has been quite inconclusive.
No evidence for a first order transition in three flavor QCD has been found so far in lattice QCD calculations \cite{Dini:2021hug}.
A recent work on this by Cuteri {\it et al.}~finds that in the continuum limit there is no first order transition for light quark masses ($m_l$) for all $N_f\le6$~\cite{Cuteri_2021}, although first order transitions can be found at non-zero $m_l$ on lattices with finite lattice spacing.

\begin{wrapfigure}{R}{0.5\textwidth}
\vspace{-0.5cm}
  \centering
    \includegraphics[width=0.99\linewidth]{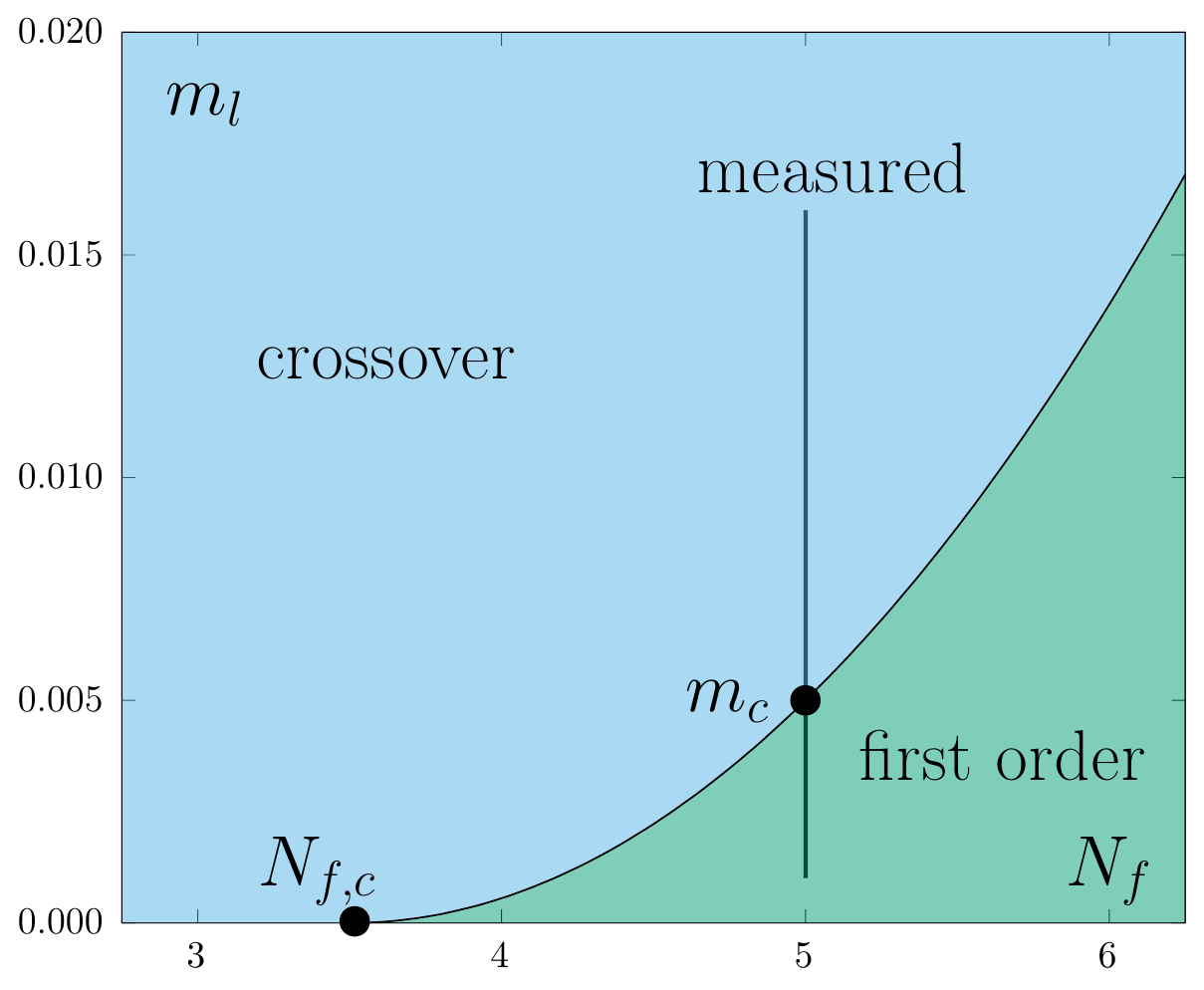}
  \caption{A sketch of a possible Columbia plot for mass-degenerate quarks in the $N_f$-$m_l$ plane, assuming a critical number of flavors $N_{f\!,c}$ between 3 and 4. Every point represents a phase boundary. The vertical line marks the measurements done in this work.}
    \label{fig:Frankfurt}
\vspace{-1.3cm}
\end{wrapfigure}

In Fig.~\ref{fig:Frankfurt} we show a sketch of the phase diagram in the $N_f$-$m_l$ plane for the situation corresponding to lattices of fixed temporal extent $N_\tau$.
We expect to find a region of first order phase transitions, which is shown in the lower right corner.
In this work, we try to find a quark mass value in this region, and while keeping $N_f$ fixed increase the mass to find the $Z(2)$ line marking the border to the crossover region.

\section{Lattice setup and observables}

The first order signal of the chiral phase transition becomes stronger with decreasing quark mass, on larger volumes and for larger number of flavors.
Unfortunately, all of these adjustments also increase the computational cost of numerical simulations using the Rational Hybrid Monte Carlo (RHMC) algorithm. 
We thus have to be quite careful with our choice of parameters.

Our calculations have been performed in five-flavor QCD ($N_f=5$) using the HISQ action with quark masses in the range $0.001\le m_l\le 0.016$ and gauge couplings $\beta=4.5-5.4$.
We used 4-dimensional lattices, $V=N_\sigma^3 N_\tau$, with temporal extent $N_\tau=6$ and spatial volumes $N^3_\sigma=16^3-24^3$. 
The partition function for these systems is given by
\begin{align}
  Z(N_f,\beta,m_l) = \int \mathcal{D} U_\mu ( \text{det} \, M [ U_\mu, m_l ] )^{N_f / 4} e^{-\mathcal{S}[U_\mu]}\; ,
\end{align}
where $M$ is the staggered fermion matrix and $S[U_\mu]$ denotes the gauge action given in terms of gauge field variables $U_\mu$.
The number of flavors, $N_f$, can easily be generalized to a continuous number, losing properties of a local quantum field theory in the process~\cite{cuteri_2018}.
The chiral condensate, which is the only observable we are going to discuss here, is defined as 
\begin{align}
\pbp = \frac{1}{4N_\sigma^3 N_\tau}\ev{\tr M^{-1}},
\end{align}

\begin{figure}[htbp]
    \centering
    \includegraphics[width=0.49\linewidth]{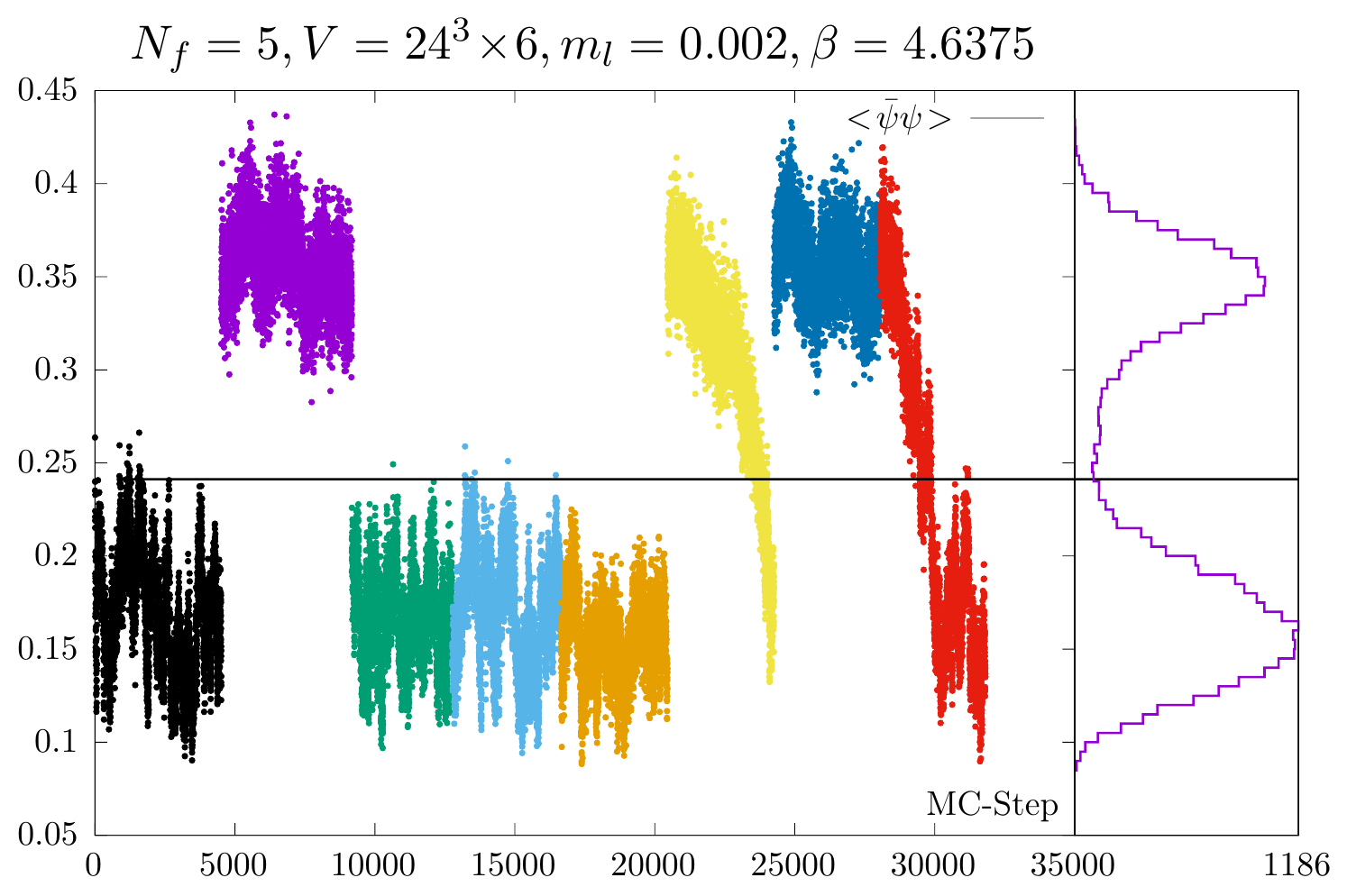}
    \includegraphics[width=0.49\linewidth]{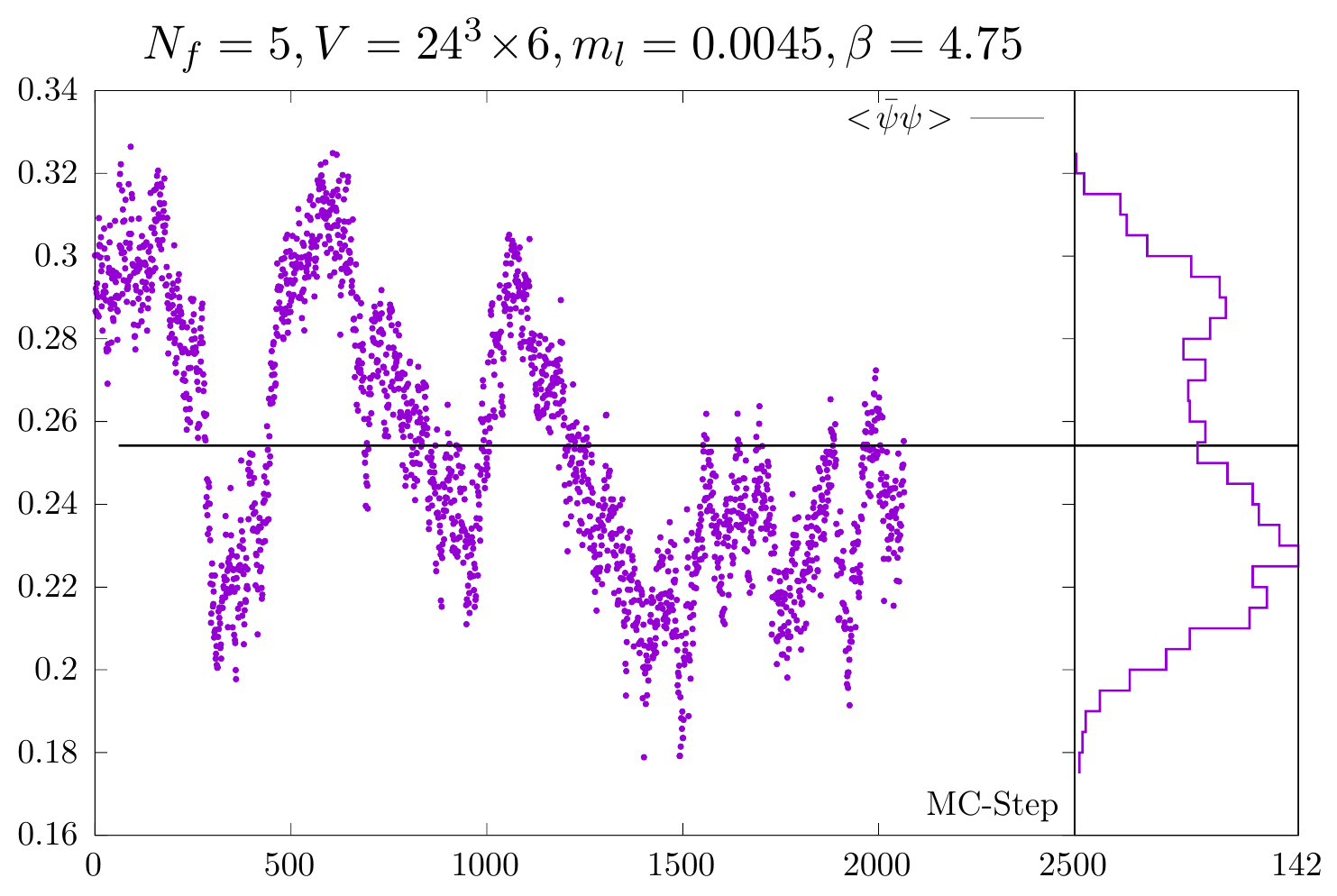}
    \caption{Time histories and respective histograms of the chiral condensate for two different masses at $N_\sigma=24$ close to $\beta_c$ in a first order region. Different colors indicate different streams.}
    \label{fig:Time_history}
\end{figure}

We calculate the chiral condensate using several independent RHMC streams generated with different starting conditions.
Results for our smallest quark mass at a value of the gauge coupling $\beta$ close to the transition temperature are shown in Figure \ref{fig:Time_history} (left).
The right hand figure shows the evolution of a single RHMC stream at a larger value of the quark mass.

In case of a first order phase transition, we expect to find two distinct phases, resulting in two peaks in the histograms of the time histories.
In the infinite volume limit ($N_\sigma\rightarrow \infty$) the minimum between the two maxima will become smaller resulting in two delta-function like peaks.
On smaller volumes the peaks broaden. 
Frequent changes from one peak region to another during the RHMC evolution will populate the region between the two maxima, up to a point where there are no more double peaks visible on small lattices.

On the other hand, a less pronounced dip in the histogram, separating the two peaks, allows for more transitions from one phase to the other, {\it i.e.}~flips in the time histories.
In Figure \ref{fig:Time_history}, we can see that indeed the signal for a first order transition weakens with increasing $m_l$, while the rate of flips increases.

\section[beta-reweighting]{\boldmath$\beta$-reweighting}

Lattice QCD calculations typically are done at a few values of the gauge coupling $\beta$.
$\beta$-reweighting~\cite{reweighting} is a popular method to interpolate lattice results.
Given measurements of the action $S$ and any observable $O$ for $R$ different $\beta_m$, it yields a continuous expectation value
\begin{align}
<\!\!O\!\!>\!(\beta) = \frac{\sum_S O(S) P(S,\beta)}{\sum_S P(S,\beta)}\; ,
\end{align}
obtained with reweighting weights $P(S,\beta)$,
\begin{align}
P(S,\beta)=\frac{   \sum^R_{n=1} N_n(S) \,\text{exp}[S \beta]    }{    \sum^R_{m=1} n_m \,\text{exp}[S \beta_m - f_m]    }\; , \quad \text{where} \quad \text{exp}[f_m] = \sum_S P(S,\beta_m) \; ,
\label{eq:reweight}
\end{align}
where $n_m$ denotes the total number of measurements made at $\beta_m$ and $N_m(S)$ is the total number of action-values in a bin $[S-\epsilon,S+\epsilon]$ around $S$ at $\beta_m$; $f_m$ is the free energy at $\beta_m$.
The weights defined in Eq.~\ref{eq:reweight} are obtained self-consistently by iterating.
This method requires a large number of measurements, performed at a large number of $\beta$-values, since $O(S)$ is obtained via the 2D-histogram of the action and the observable we want to reweight.
Moreover, the action histograms obtained at the different $\beta_m$ need to have a sufficiently large overlap.

The method can be extended to reweight a probability distribution of any observable by reweighting each bin of the discretized distribution individually. 
This approach is thus limited to data sets discretized in a set of bins and 
only interpolates in $\beta$-direction.

\begin{figure}[htbp]
    \centering
    \includegraphics[width=0.49\linewidth]{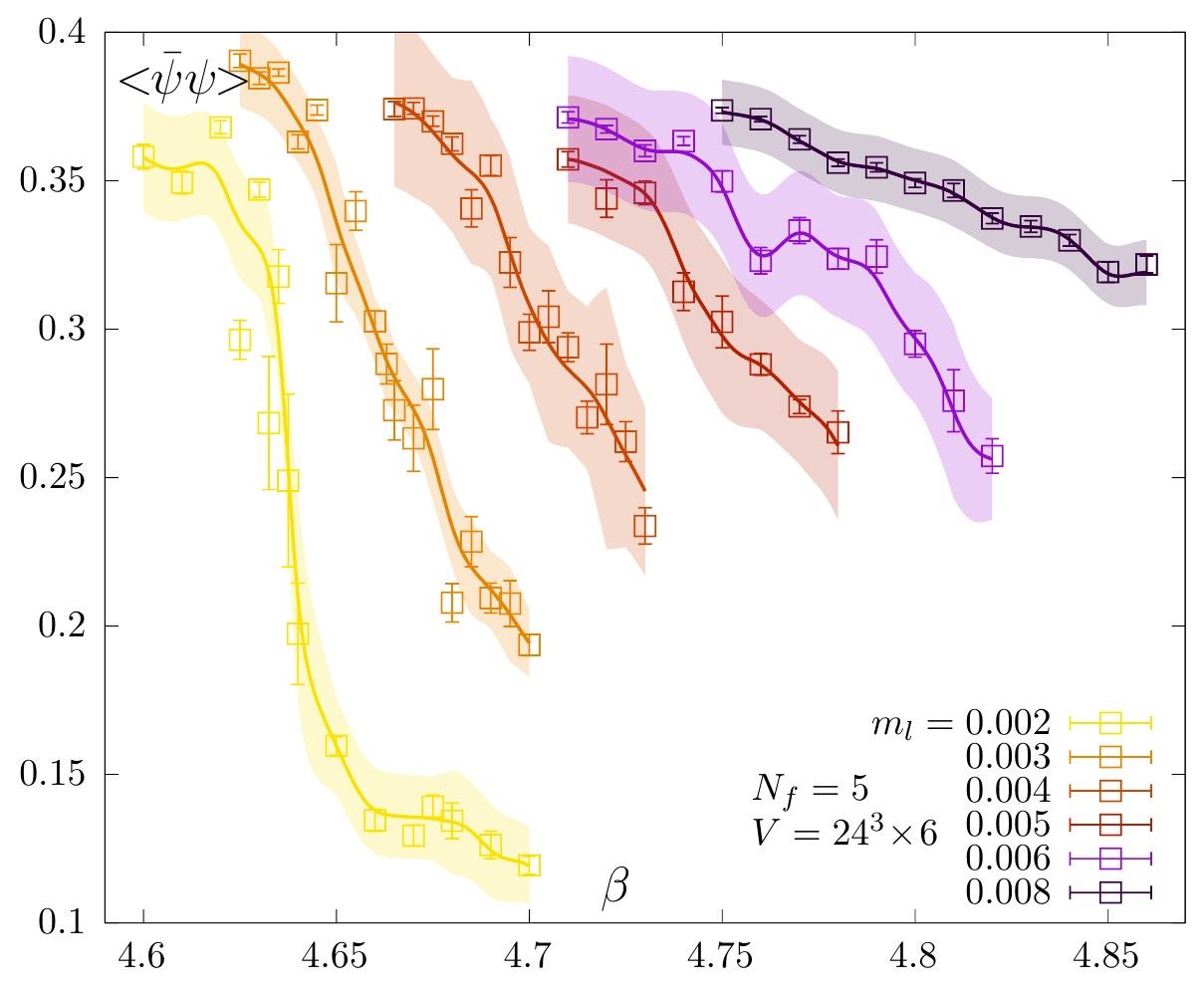}   \includegraphics[width=0.49\linewidth]{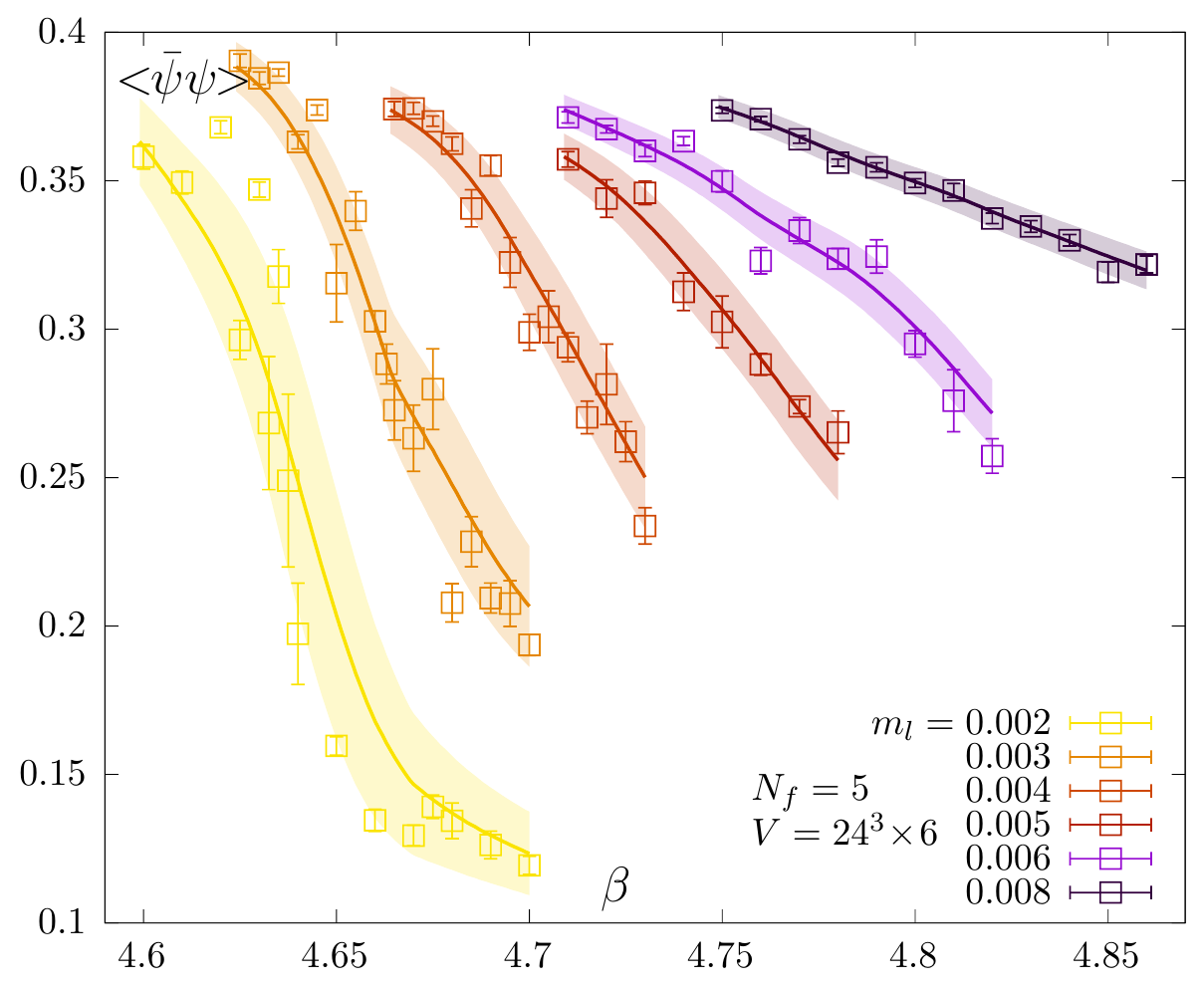}
    \caption{Comparison of $\beta$-reweighting (left) and ML-reweighting (right). Data points show results obtained from  RHMC calculations in 5-flavor QCD, while the curves are obtained from the $\beta$- and ML-reweighting, respectively.}
    \label{fig:reweight_compare}
\end{figure}

In Figure \ref{fig:reweight_compare}~(left) $\beta$-reweighted data for the chiral condensate are shown.
The reweighting is done for the entire set of histograms at each mass, but only the expectation values are shown, to obtain a compact plot.
While this yields reasonable results for the lowest masses, for larger $m_l$, especially $m_l=0.006$, the $\beta$-reweighting obviously is over-fitting.

\section{ML model}

Normalizing flows are state-of-the-art tools for modeling probability distributions in physical systems.
We use a MAF (Masked Autoregressive Flow)~\cite{MAF} model with eight MADE (Masked Autoencoder for Distribution Estimation)~\cite{MADE} blocks.
MADE networks have been especially designed to factorize a joint probability distribution into a product of conditional probabilities.
Using less than eight MADE blocks caused problems with fitting the double peaks, however, for fits in the crossover region a fewer number of MADE should be sufficient.
Compared to the classical reweighting, this method has the advantage of allowing to interpolate in any parameter.
In particular, there is no need for overlapping distributions of the action density and the method is able to process continuous data.
However, in order to visualize the learned probability distribution, we need to draw a large number of samples from our model to fill a two dimensional histogram.

In the end, the model learns to transform a 2D-Gaussian distribution to ``measurements'' of $(\bar{\psi}\psi,S)$, conditioned on the continuous parameters $(N_\sigma, m_l, \beta)$.
To avoid overfitting, we have introduced penalty terms in the loss function, based on the L1- and L2-norms of the parameters of the network, known as regularization.
The regularization is applied on a per-layer basis and the coefficients in front of the regularization terms have been chosen as $l_1=l_2=0.0001$.
The training took approximately 4h on a
NVIDIA V100 GPU.
Evaluating the model was done for all integer $N_\sigma \in [16,24]$, $\beta \in [4.5,5.4]$ in steps of $0.001$ and $m_l \in [0.001,0.006]$ in steps of $0.001$ and for the larger masses $m_l \in [0.008,0.016]$ in steps of $0.002$.
Inference took approximately 30sec per 1,000,000 measurements at each parameter combination.
This allows us to fit our entire data set (as shown in Table \ref{tab:stats}) with a single function $p(\bar\psi\psi,S \, | \, N_\sigma, m_l, \beta)$, in contrast to the $\beta$-reweighting, where we would need to do independent reweighting for each mass and volume. 

In Figure \ref{fig:reweight_compare}~(right), the ML-reweighted data are shown.
While the interpolation appears to be slightly under fitting for $m_l=0.002$, in the grand picture we achieve a good fit.
Compared to the $\beta$-reweighting, it is intuitive that we get a better fit, since now the data points support each other also in $m_l$ and $N_\sigma$-direction and not only in $\beta$.
Of course we could have fitted only a 1D distribution to the chiral condensate. However, we included the action as well to stabilize the fit and enable easy comparison with the $\beta$-reweighting approach.

\begin{figure}[htbp]
    \centering
    \includegraphics[width=0.49\linewidth]{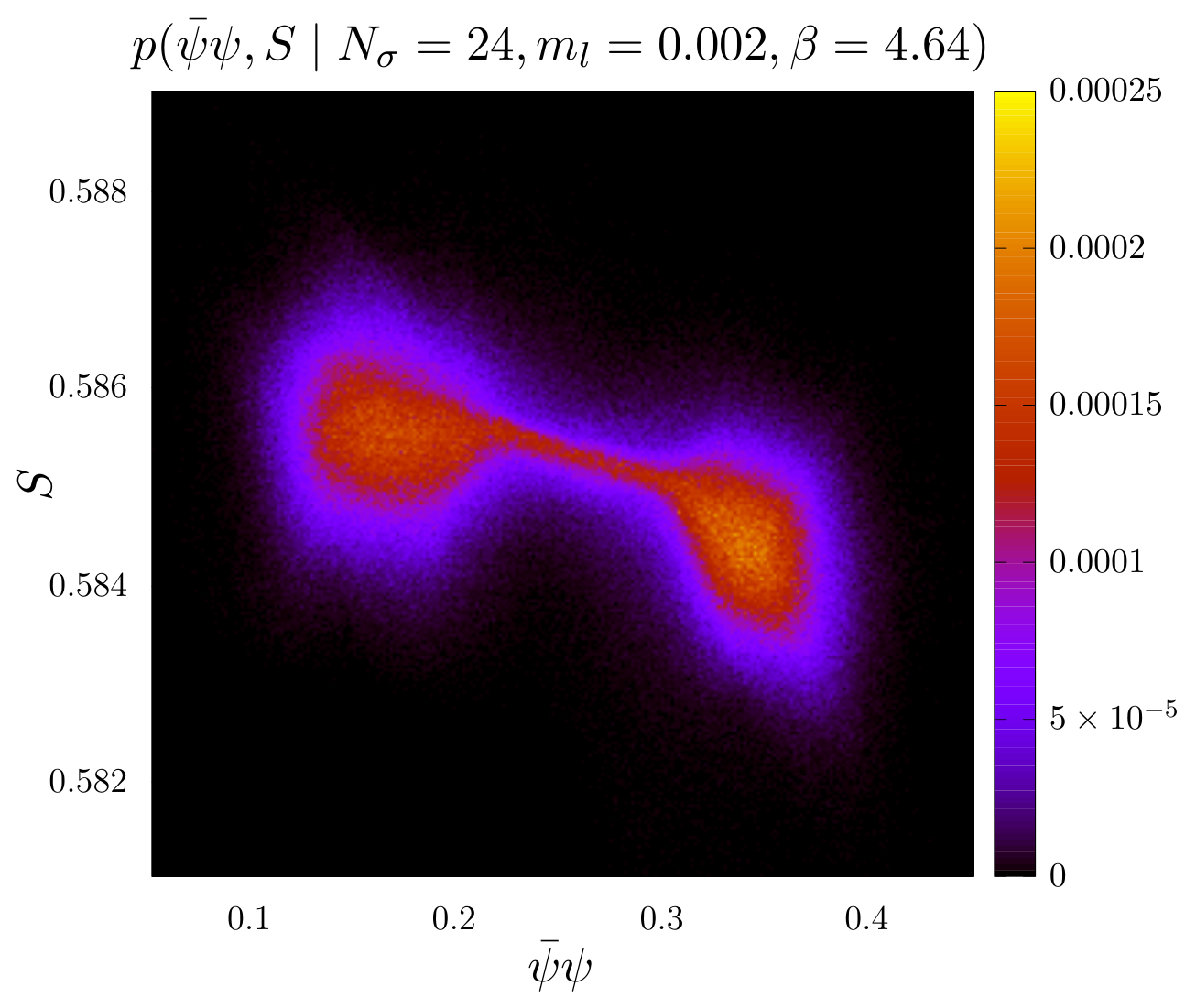}
    \includegraphics[width=0.49\linewidth]{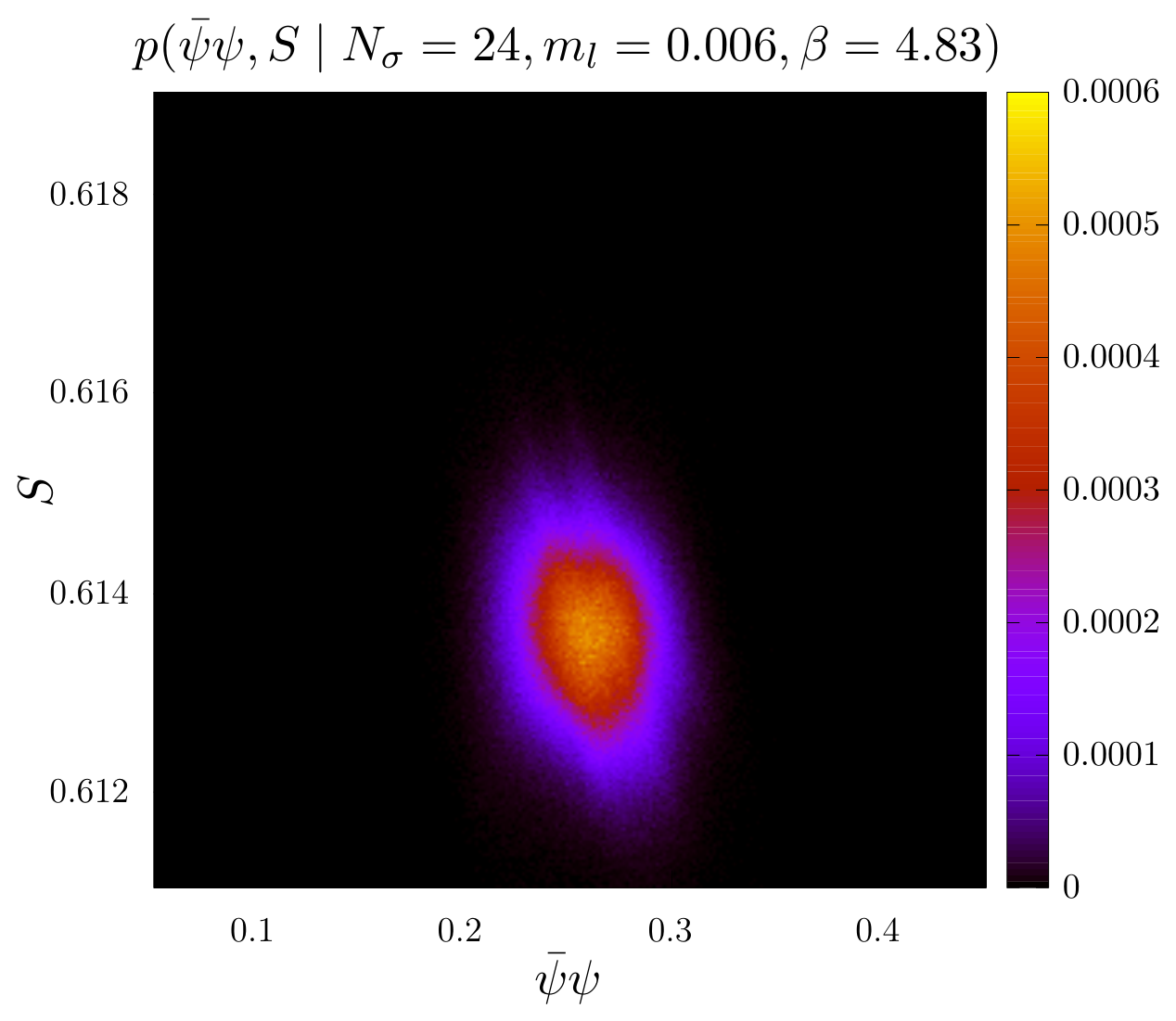}
    \caption{Density plot in the $\bar\psi\psi$-$S$
    plane for 1,000,000 evaluations of the model. Shown are results for a small (left) and large (right) quark mass,
    corresponding to the first order and crossover regions, respectively.}
    \label{fig:2D_chi_S}
\end{figure}

\section{Results}

\begin{table}
\centering
\begin{tabular}{c|ccccccc}
$ N_\sigma$ & 0.001 & 0.002 & 0.003 & 0.0035 & 0.004 & 0.0045 & 0.005  \\ \hline
16 & 17201 & 18887 & 11526 & 0 & 18866 & 0 & 0   \\
24 & 5294 & 83177 & 149885 & 25028 & 30571 & 19332 & 19352  \\ \hline \hline
$ N_\sigma$ & 0.006 & 0.008 & 0.010 & 0.012 & 0.014 & 0.016 & \\ \hline
16 & 61382 & 61220 & 61456 & 61456 & 61256 & 61256 & \\
24 & 42762 & 82061 & 65140 & 13380 & 36574 & 36499 & \\
\end{tabular}
\caption{Total number of RHMC measurements for each volume and mass, summed over all available $\beta$ values, corresponding to approximately 300,000 GPUh on a NVIDIA V100.}
\label{tab:stats}
\end{table}

In Figure~\ref{fig:2D_chi_S} we show directly the model output in form of a 2D histogram (contour plot) in the $\bar{\psi}\psi$-$S$ plane.
The two distinct phases connected by a small band are clearly visible in the left hand figure, which shows results for a small quark mass, while only a single phase seems to be present in the right hand figure, which is for a large quark mass value.

\begin{figure}[htbp]	
	\centering
    \includegraphics[width=0.49\linewidth]{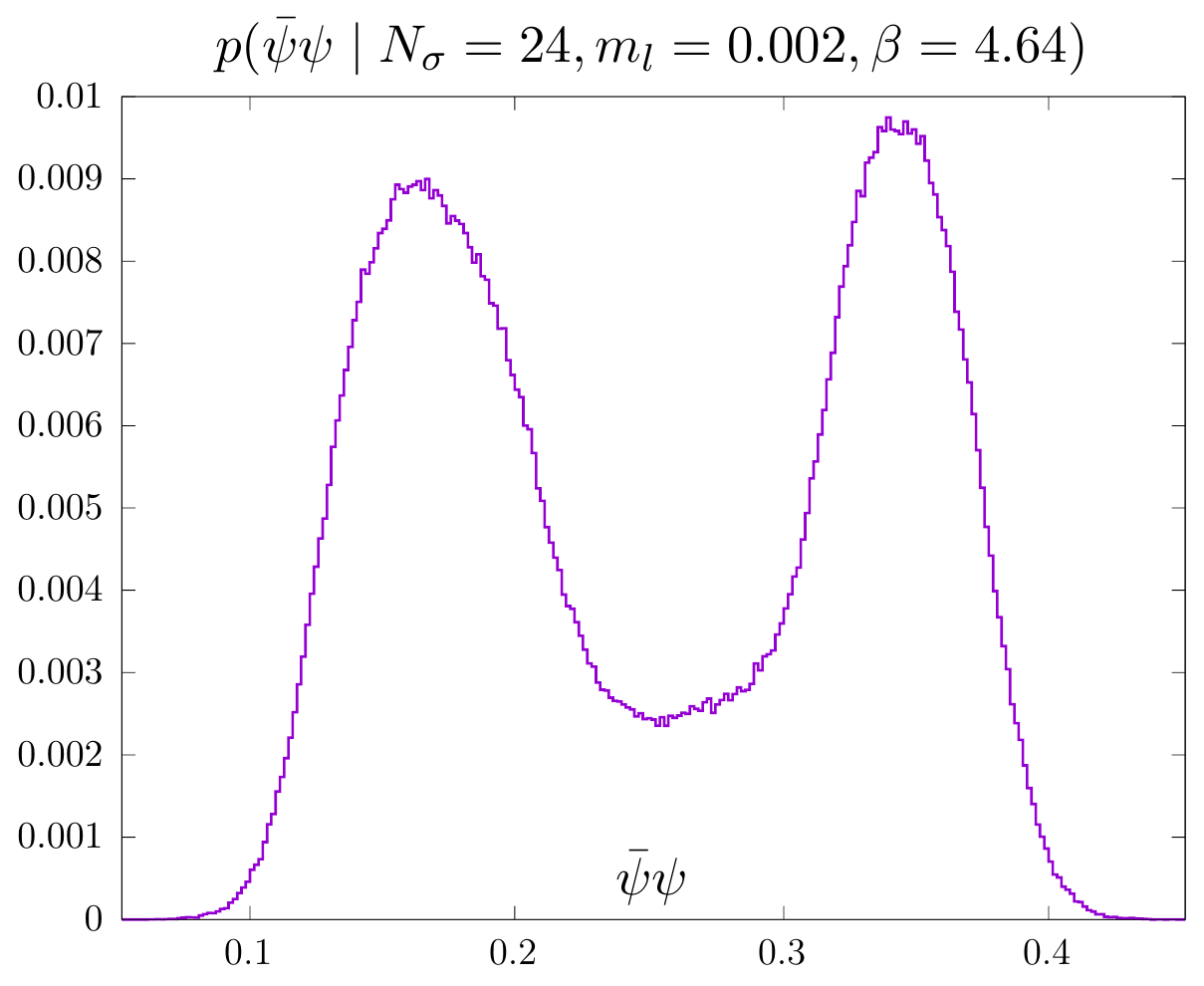}
    \includegraphics[width=0.49\linewidth]{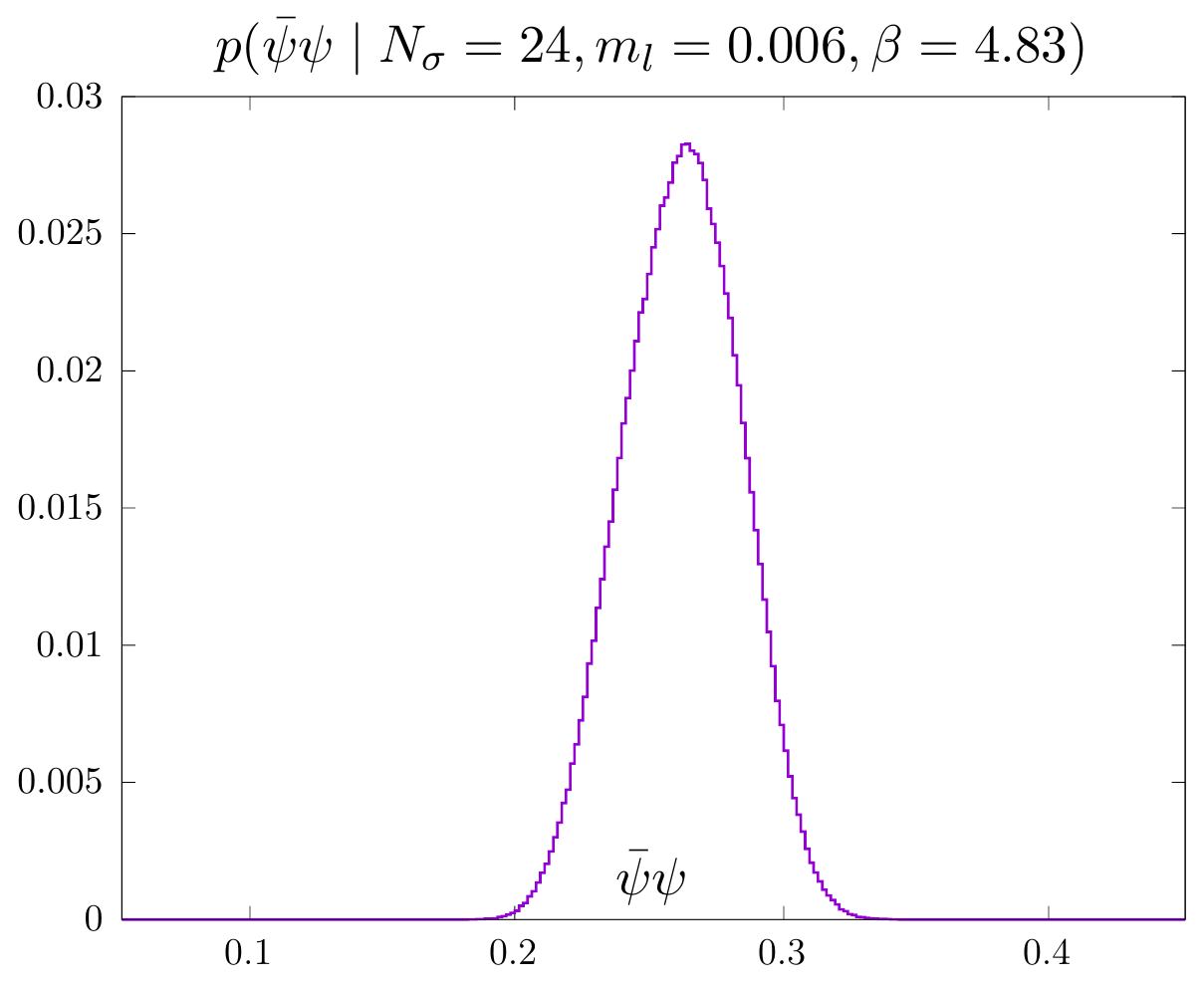}
    \caption{$p(\bar\psi\psi$) with 250 histogram bins and 1,000,000 evaluations of the ML-model for a light quark mass in the first order region (left) and a heavier mass in the crossover region (right).}
    \label{fig:p_chi}
\end{figure}

\begin{figure}[htbp]
    \centering
    \includegraphics[width=0.49\linewidth]{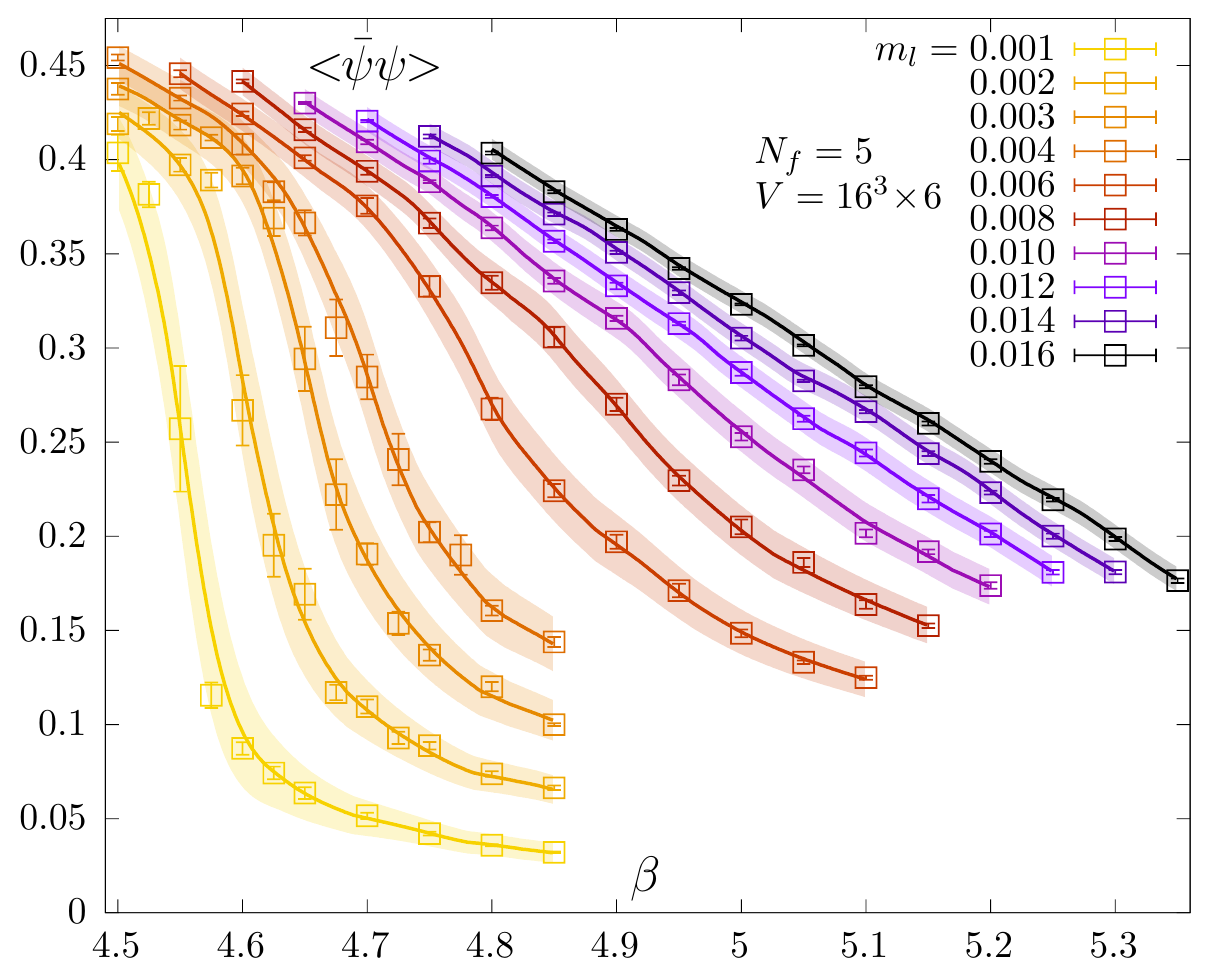}
    \includegraphics[width=0.49\linewidth]{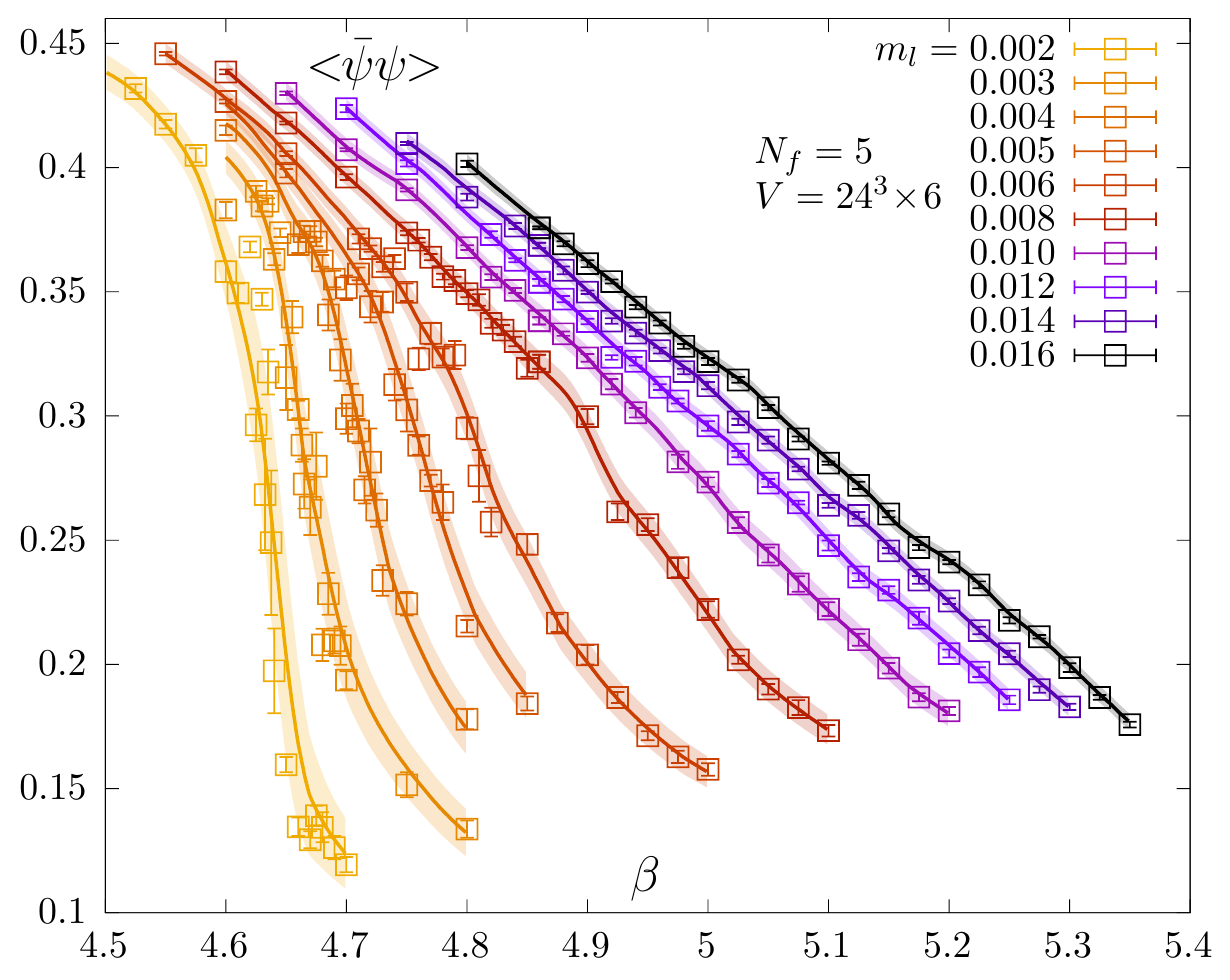}
    \caption{$\pbp$ for $N_\sigma=24$ (left) and $N_\sigma=16$ (right). The data points represent the RHMC measurements, while the curves are taken from the interpolation generated by the ML model.}
    \label{fig:chi_smooth}
\end{figure}

Since we are mainly interested in the chiral condensate, we project the 2D histogram on the chiral condensate axis and look at 1D histograms as shown in Figure~\ref{fig:p_chi}.
The mean values of these distributions can again be compared to the measured points, but this time for the whole data set, as shown in Figure~\ref{fig:chi_smooth}.
Again, we point out, that the entire data set is described by a single fit (with about 10,000 parameters).

It is also possible to extract $\pbp$ as well as the 1D histograms of $\bar{\psi}\psi$ in a fine sampled $m_l$-$\beta$ plane.
This allows to determine the quark mass dependence of the double peaks seen in Figure~\ref{fig:p_chi}.
They signal the occurrence of a first order phase transition, with the right hand peak corresponding to the end of the symmetry broken phase, the left hand peak corresponding to the symmetry restored phase and the region between
the peaks being the mixed phase.
The corresponding phase  

\begin{wrapfigure}{L}{0.7\textwidth}
    \centering
    \includegraphics[width=0.95\linewidth]{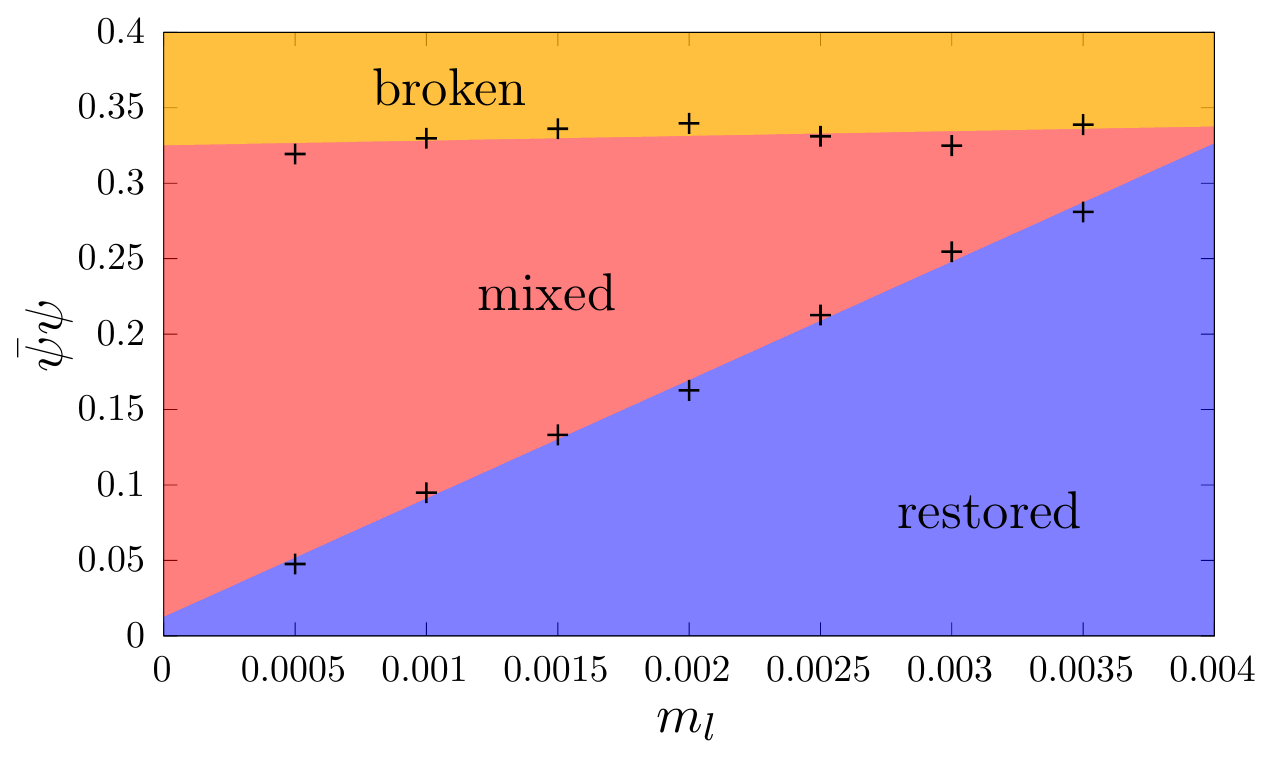}
    \caption{Phase diagram of 5-flavor QCD on lattices with fixed temporal extent, $N_\tau=6$ in the $m_l$-$\beta$ plane.}
    \label{fig:phase_diagram}
\vspace{-0.4cm}
\end{wrapfigure}
\noindent
diagram in the $m_l$-$\beta$ plane is shown in Figure~\ref{fig:phase_diagram}.
It suggests that the first order region ends in a second order end point
at about $m_l^c \simeq 0.0045$.
Clearly, as the gap between the peaks at low and high $\beta$ becomes smaller larger lattices will be needed to resolve these two peaks and establish a gap between them.
In the next section we will discuss a ML based approach to locate this end point.

\section{EOS-meter}

\begin{figure}[htbp]
    \centering
    \includegraphics[width=0.49\linewidth]{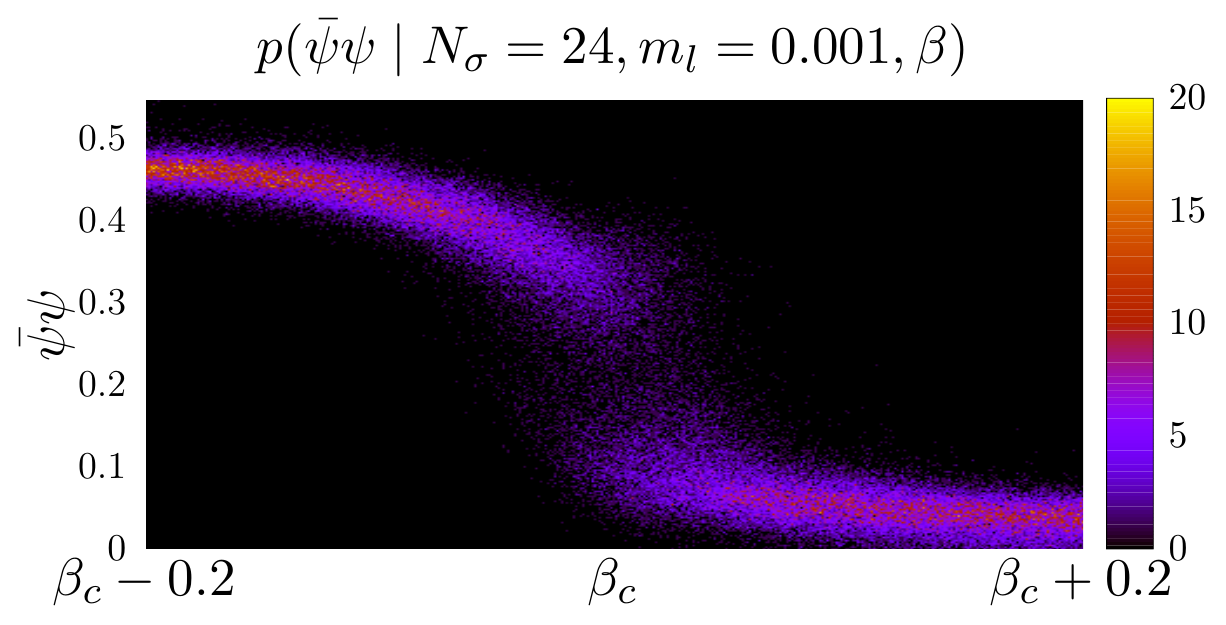}
    \includegraphics[width=0.49\linewidth]{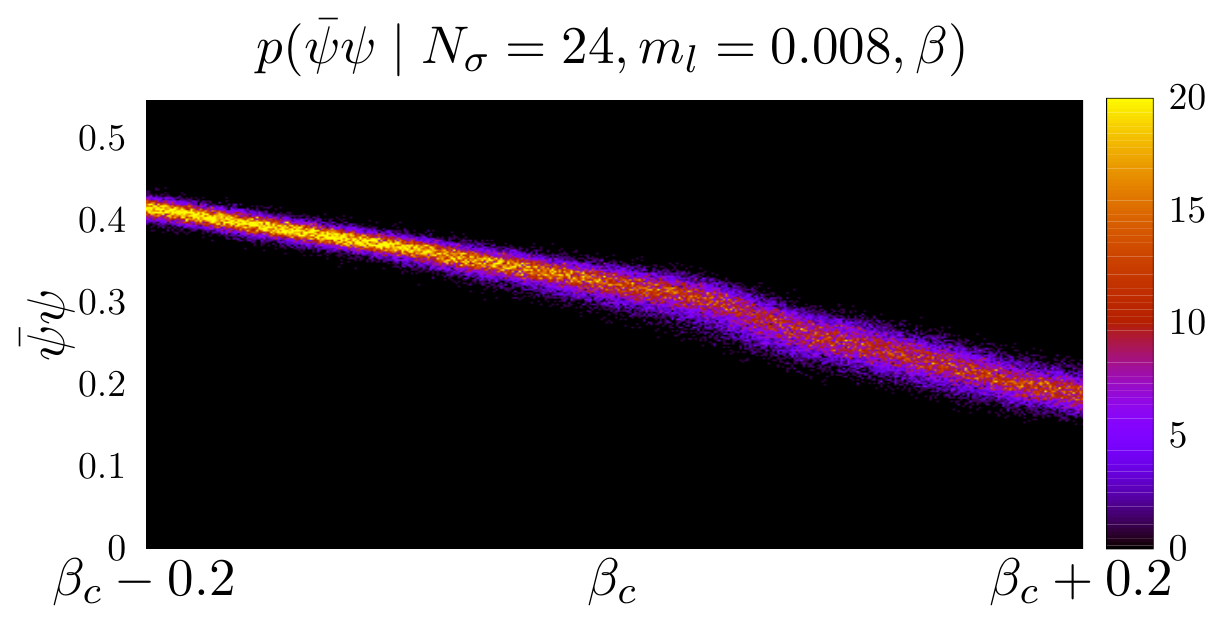}
    \caption{Probability density plots used to train the EOS-meter. Each column of pixels corresponds to 1000 evaluations of the model. $\beta_c(m_l)$ does not need to be known exactly, as long as $\beta_c$ is within the plot range.}
    \label{fig:2D_p_chi}
\end{figure}

Petersen {\it et al.}~have introduced the idea of using an ML image recognition approach to classify phase transitions \cite{EOSmeter}.
They used a convolutional neural network (CNN) model to classify data sets obtained in heavy-ion collision.
The resulting density plots they called an Equation-of-State-meter.
Recently, the transformer model \cite{attention}, a model solely based on attention mechanisms, has been shown to outperform recurrent or convolutional neural networks in translation tasks.
Transformers are expected to generalize well to other tasks, including image recognition applications.
Since no CNNs are used, information on pixel positions must be added artificially via a so-called positional encoding.
Here we have used a vision transformer based approach on density plots as shown in Figure~\ref{fig:2D_p_chi}.
We have labeled the histograms of the smallest masses, where a clear gap was visible as ``first order'' while the histograms of the largest masses were labeled as ``crossover''.

``Firstorderness'' and ``crossoverness'' are implemented as categories in one-hot-encoding.
During training, random translation in $\beta$-direction was applied, since it makes the trained model more independent of our estimate of $\beta_c$.
Since Dropout can be used as a Bayesian Approximation to the model uncertainty~\cite{dropout_err}, we can show error bars on the determined ``firstorderness''.
We also tried a more traditional CNN approach, which however resulted in less sharp transitions.

\begin{wrapfigure}{L}{0.5\textwidth}
    \centering
    \includegraphics[width=0.94\linewidth]{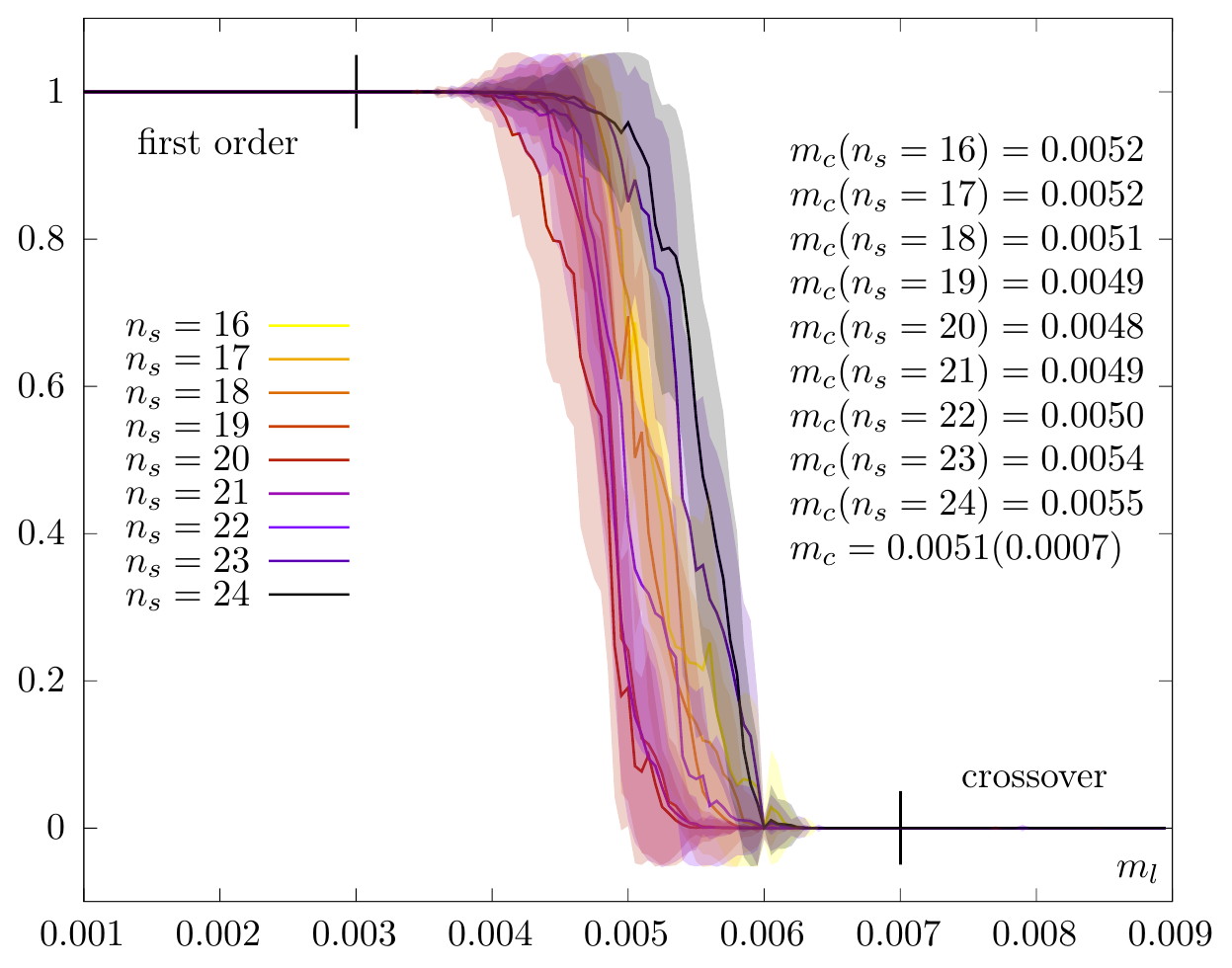}
    \caption{An EOS-meter, the ``firstorderness'' of the chiral phase transition plotted versus $m_l$ for different volumes. The vertical bars mark the cuts between training and testing data.}
    \label{fig:EOS-meter}
	\vspace{-1.17cm}
\end{wrapfigure}
The resulting EOS-meter is shown in Figure~\ref{fig:EOS-meter}.
We want to remark that even though we can see some fluctuations, $m_c$ should not and does not depend on $N_\sigma$. 
The critical masses marking the borders between first order and crossover regions were extracted via logistic fits to the ``firstorderness''
\begin{align}
   f(m_l)=\frac{1}{1+e^{k(m_l-m_c)}} .
\end{align}
From these fits, we can extract $m_c=0.005(1)$.

\section{Conclusions}

Normalizing flows appear to be a performant alternative to $\beta$-reweighting.
We achieve a good model of the $(\bar\psi\psi, S)$ distribution in $(N_\sigma, m_l, \beta)$ for our entire data range.
The model can be used to extract a fine enough sampling in the parameter range to train an EOS-meter able to extract the ``firstorderness'' of the chiral phase transition for $N_f=5$, making it possible to identify a critical mass $m_c \approx 0.005(1)$ which marks the border between the first order and crossover regions.

In order to use this model to extract the phase diagram of QCD with $N_f$ flavors in the continuum limit we need to use larger $N_\tau$ values.
An extension of the parameter set to $(N_f, N_\sigma, N_\tau, m_l, \beta)$ is expected to be possible, but is going to require a large amount of training data.
\section*{Acknowledgments}
\label{sec:acknowledge}
This work was supported in part by the Deutsche Forschungsgemeinschaft (DFG) through the grant 315477589-TRR 211 and "NFDI 39/1" for the PUNCH4NFDI consortium and the grant 
EU H2020-MSCA-ITN-2018-813942 (EuroPLEx) of the European Union.
All calculations have been performed on the Bielefeld University GPU cluster and we thank members of the HPC.NRW team for their support.
We would also like to thank the ERuM-Data-Hub workshop ``Conceptual Advances in Deep Learning for Research on Universe and Matter'' for sharing many ideas on the choice of ML models.

\section*{Software}

We used SIMULATeQCD~\cite{simulateqcd} for the RHMC calculations, and Keras~\cite{keras} and Tensorflow probability~\cite{Tensorflow} to implement the ML models.

\bibliographystyle{JHEP} 
\bibliography{bibliography}

\end{document}